\documentstyle[prl,aps,floats,epsf]{revtex}\def\narrowtext{}\tighten\twocolumn
\input epsf.sty
\begin{document}
\draft

\title{Cu Nuclear Quadrupole Resonance Study of the Spin-Peierls 
Compound Cu$_{1-x}$Mg$_x$GeO$_3$: A Possibility of Precursory Dimerization}
\author{Y. Itoh$^{1,2}$, T. Machi$^1$, N. Koshizuka$^1$, T. Masuda$^3$, 
 and K. Uchinokura$^3$}

\address{$^1$Superconductivity Research Laboratory,
 International Superconductivity Technology Center,\\
1-10-13 Shinonome, Koto-ku Tokyo 135-0062, Japan \\
$^2$Japan Science and Technology Corporation, Japan\\
$^3$Department of Advanced Materials Science, The University 
of Tokyo, 7-3-1 Hongo, Bunkyo-ku, Tokyo 113-8656, Japan}


\address{\begin{minipage}[t]{6.0in} %
\begin{abstract}%
We report on a zero-field $^{63}$Cu nuclear quadrupole resonance 
(NQR) study of nonmagnetic Mg impurity substituted Cu$_{1-x}$Mg$_x$GeO$_3$ 
(single crystals; the spin-Peierls transition temperature $T_{sp}{\sim}$14, 
13.5, and 11 K for $x$=0, 0.0043, and 0.020) in a temperature 
range from 4.2 K to 250 K. We found that below $T^*{\sim}$77 
K, Cu NQR spectra are broadened and nonexponential Cu nuclear 
spin-lattice relaxation increases for undoped and more remarkably 
for Mg-doped samples. The results indicate that random lattice 
distortion and impurity-induced spins appear below $T^*$, 
which we associate with a precursor of the spin-Peierls transition. 
Conventional magnetic critical slowing down does not appear down 
to 4.2 K below $T_{sp}$. 
\typeout{polish abstract} %
\end{abstract}
\end{minipage}} %

\maketitle %
\narrowtext

The discovery of the first inorganic spin-Peierls compound CuGeO$_3$ 
(the transition temperature $T_{sp}{\sim}$14 K)~\cite{Hase} and subsequent reports on unprecedented impurity
effects~\cite{Hase1,Regnault,Masuda,Koide,Kojima,Martin} have renewed the interests of a quasi-one-dimensional spin $S$=1/2 Heisenberg
antiferromagnet coupled to phonons. No soft mode of phonon at $T_{sp}$ is one of the characteristics of CuGeO$_3$. 
An appreciable inter-chain exchange interaction~\cite{Nishi}, lattice 
and phonon anomaly perpendicular to the chain~\cite{Braden,Lorenzo} are different 
from an ordinary spin-Peierls system or conventional theoretical 
result~\cite{Cross}. A pseudogap in the magnetic excitation spectrum 
below 20 K observed by inelastic neutron scattering~\cite{Regnault1} and 
a local dimerization until at least 40 K observed by diffusive 
X-ray scattering~\cite{Pouget} resemble a pseudogap of the electronic 
Peierls materials~\cite{Lee}. In contrast to conventional competition 
between N\'{e}el ordering and lattice dimerization, the impurity 
substitution (Zn, Mg, Ni, or Si) for Cu or Ge induces a dimerized 
antiferromagntic ordering state~\cite{Regnault,Masuda,Koide,Kojima}, where a spin-wave mode 
coexists in the spin-Peierls gap~\cite{Martin}. The coexistence at $T$=0 
K is understood within the framework of the phase Hamiltoninan 
~\cite{Fukuyama}. 

Nuclear quadrupole resonance (NQR) and nuclear magnetic resonance 
(NMR) are unique and powerful techniques to study low frequency 
dynamics and local spin fluctuations in space. The intensive 
studies using Cu NQR and NMR techniques have revealed many aspects 
of CuGeO$_3$~\cite{Kikuchi,Mitoh,Fagot,Fagot1}. The spin gap opening at $T_{sp}$ was evidenced by an abrupt decrease of the Cu
nuclear spin-lattice  relaxation rate 1/$T_1$ without any appreciable change 
of Cu NQR spectrum~\cite{Kikuchi}, although a singlet-triplet excitation, 
a spin gap and ion displacement were directly confirmed by neutron 
scattering~\cite{Fujita,Hirota}. The above $T_{sp}$ spin dynamics, 
the Cu 1/$T_1$, is understood by a quasi-one-dimensional $S$=1/2 
antiferromagnetic correlation without spin-phonon coupling~\cite{Kikuchi,Mitoh1,Yu}.
 To our knowledge, however, there are no reports of Cu 
NQR spectrum far above $T_{sp}$ for CuGeO$_3$ nor of 
impurity effects on the low frequency spin dynamics, after the 
work on Zn doping~\cite{Mitoh2}. 

In this Letter, we report the high temperature measurement of 
Cu NQR spectrum for undoped CuGeO$_3$ and the Cu NQR study 
of nonmagnetic Mg impurity substitution effect on Cu$_{1-x}$Mg$_x$GeO$_3$ 
(single crystals; $T_{sp}\sim$14, 13.5, and 11 
K for $x$=0, 0.0043, and 0.020) in a wide temperature 
range of $T$=4.2-250 K. We found that the broadening 
of Cu NQR spectra and the increase of nonexponential Cu nuclear 
spin-lattice relaxation occur below about 77 K much higher than $T_{sp}$, 
which suggest a precursor of the spin-Peierls transition. We 
did not observe critical divergence of 1/$T_1$ for $x$=0.020 
down to 4.2 K, although the magnetic ordering occurs at about 
2.5 K~\cite{Masuda}. 

The single crystals grown by a floating-zone method are well 
characterized in Ref.~\cite{Masuda}. Zero-field $^{63}$Cu NQR spin-echo 
measurements were carried out with a coherent-type pulsed spectrometer. 
NQR frequency spectra with quadrature detection were measured 
by integration of the $^{63}$Cu nuclear spin-echoes as the frequency 
was changed point by point. Nuclear spin-lattice relaxation was 
measured by an inversion recovery spin-echo technique, where 
the $^{63}$Cu nuclear spin-echo amplitude $M(t$) 
was recorded as a function of time interval $t$, between 
an inversion $\pi$-pulse and a $\pi$/2-pulse 
($\pi-t-\pi/2-\pi$-echo). 

Figure 1 shows $^{63}$Cu NQR spectra for undoped $x$=0 
(a) and for Mg doping of $x$=0.020 (b) in the temperature 
range of $T$=4.2-250 K. The observed Cu NQR spectra 
for Mg doping are nearly symmetrically broadened, not of Gaussian 
nor of Lorentzian type but rather have a triangle-shaped line 
profile for $x$=0.020 at 4.2 K. Implication of the characteristic 
line shape is not clear. In general, the Cu NQR frequency $\nu$ 
is given by $\nu=(e^2qQ/2h)\sqrt{1+\eta^2/3}$, where
$eq$ is the maximum component of the electric field gradient tensor at the nuclear site,
$Q$ is the  nuclear quadrupole moment, and $\eta$ 
$(0\leq\eta\leq1)$ is an asymmetry factor~\cite{Cohen}. The muon spin
relaxation measurements have not detected any static internal magnetic field for Zn-doped samples above about 4 K
~\cite{Kojima}. Thus, it
is likely that the random  distribution of the electric field gradient ($eq$ and $\eta$) is the origin of broadening. 

\begin{figure}
\epsfxsize=3.8in
\epsfbox{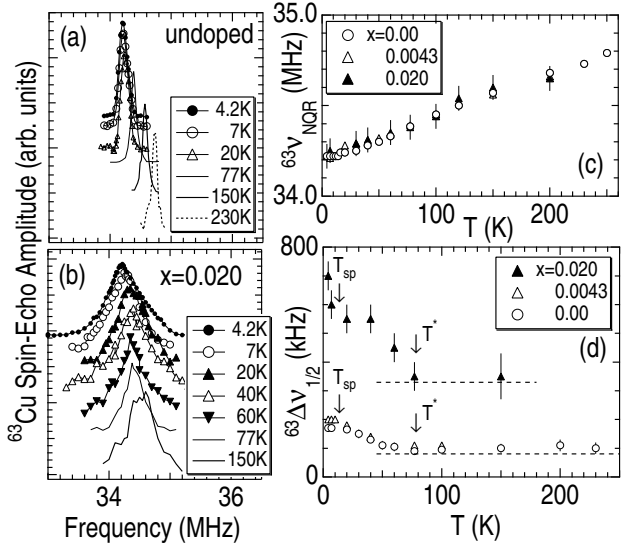}
\vspace{0cm}
\caption{
The temperature dependence of $^{63}$Cu NQR frequency spectrum for $x$=0 (a) and for $x$=0.020 (b). 
The temperature dependence of the peak frequency $^{63}\nu$ (c) and of the linewidth (full-width at half maximum)
$^{63} \Delta \nu _{1/2}$ (d) for $x$=0, 0.0043. and 0.020. The dashed lines in (d) are
guides to the eye.
}
\label{NQR}
\end{figure}

Figure 1(c) shows the temperature dependence of the peak 
frequency $^{63}\nu$ for $x$=0, 0.0043, and 
0.020. The observed linear temperature dependence of $^{63}\nu$, 
nearly independent of Mg content, is similar to that of the cuprate 
mono-oxide CuO~\cite{Yitoh}. Figure 1(d) shows the temperature dependence 
of the linewidth defined as full-width at half maximum $^{63}\Delta\nu_{1/2}$. 
The linewidth as well as the peak frequency does not show any 
appreciable change at $T_{sp}$, in agreement with the 
previous reports on CuGeO$_3$~\cite{Kikuchi,Mitoh1,Gippius}. The linewidth of 
the $x$=0.020 sample is about 3 times larger than that 
of $x$=0 at $T > T^*$. In Fig. 
1(d), we obtain a significant result that $^{63}\Delta\nu_{1/2}$ 
 increases rapidly below about 77 K $>> T_{sp}$ 
(denoted as $T^*$) for all samples including CuGeO$_3$. 
For each $x$, $^{63}\Delta\nu_{1/2}$ at 4.2 
K is about 2 times larger than that above $T^*$.

NQR is a measure of deviation of charge distribution from 
cubic symmetry around the nuclear site, being quite sensitive 
to crystal imperfections. The observed nearly symmetric line 
shape implies a random distribution of local charge~\cite{Cohen}. If 
the origin of $^{63}\Delta\nu_{1/2}$ is a static distribution 
of lattice distortion around crystal imperfection, $^{63}\Delta\nu_{1/2}$ would decrease as the temperature is
decreased so as to scale  with the temperature dependence of $^{63}\nu$. However, 
the actual $^{63}\Delta\nu_{1/2}$ increases below $T^*$. 
Thus, the inhomogeneity of lattice distortion must depend on 
temperature and must increase rapidly below $T^*$. The 
pretransitional lattice fluctuations above $T_{sp}$ observed 
by the diffraction experiment~\cite{Pouget}, which are explained by the 
random-phase approximation calculation~\cite{Gros}, may be closely related 
with the increase of $^{63}\Delta\nu_{1/2}$. According 
to the recent quantum Monte Carlo simulation~\cite{Onishi}, a precursory 
dimerization takes place near the edges far above $T_{sp}$. $T^*$ 
corresponds to the onset of preformed dimer
bonds. In the  soliton picture~\cite{Khomskii}, $T_{sp}$ is an order-to-disorder 
transition temperature of locally dimerized segments. $T^*$ 
 may correspond to the onset of development of the interchain 
correlation between the soliton and antisoliton. 

\begin{figure}
\epsfxsize=3.0in
\epsfbox{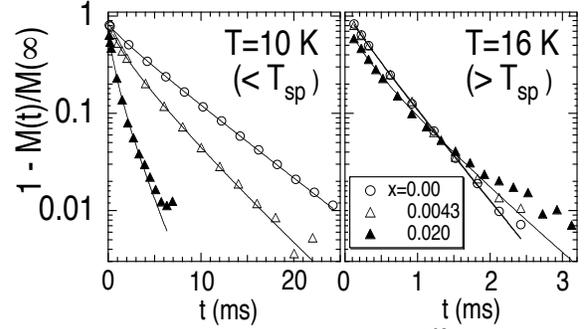}
\vspace{0cm}
\caption{
Mg-doping dependence of $^{63}$Cu recovery curves 1-$M(t)/M(\infty)$ at 10 K (a) and at 16 K (b) for $x$=0, 0.0043, and 0.020.  The solid
curves are the least-squares fitting results using  eq. (1).
}
\label{Recovery}
\end{figure} 

Figure 2 shows Mg-doping effect on $^{63}$Cu recovery curve $p(
t)\equiv1-M(t)/M(\infty)$ of the $^{63}$Cu nuclear
magnetization $M(t)$ at 10 K ($< T_{sp}$) (a) and at 16 K ($> T_{sp}$) 
(b). The recovery curve changes from a single exponential function to nonexponential one as Mg is substituted and
as the temperature  is decreased. To account for the nonexponential function, we  assume a minimal model, which
consists of a host homogeneous relaxation process and of a single inhomogeneous relaxation one. 
The solid curves are the least-squares fitting results using 
the following equation~\cite{McHenry}, 
 
 $p(t)=p(0)$exp$[-(t/T_1)_{NQR}-\sqrt{t/\tau_1}]$.      (1)

The fit parameters are $p$(0), ($T_1)_{NQR}$~\cite{DefT1} and $\tau_1$. $p(0)$ is a fraction of an initially inverted
magnetization, and ($T_1)_{NQR}$ is the nuclear spin-lattice relaxation
time due to the host  Cu spin fluctuations. $\tau_1$ is an impurity-induced 
nuclear spin-lattice relaxation time, which is originally termed 
a longitudinal direct dipole relaxation time, because the second 
term of eq. (1) is derived from a random $T_1$ process 
of 1/$T_1(r)=C/r^6$ ($C$ is a constant, and $r$ is a distance
between an impurity-induced  spin $S$ and a Cu nuclear spin $I$)
through  a direct dipole coupling ${\propto}I_{\pm}S_z/r^3$ ($S_z$ is the $z$-component of $S$, and
$I_{\pm}$ is a raising or lowering operator of $I$). The randomly distributed impurity-induced spins  yield
the stretched exponential function of eq. (1). The original  Mg ion does not carry spin 1/2. Atomic defects or Mg
ions cut  chains into segments. The existence of spatially extended staggered 
moment induced by an edge or an impurity has been pointed out 
for a finite or a semi-infinite chain~\cite{Onishi,Khomskii,Eggert}. The assumption 
of impurity-induced spins could be only a working hypothesis 
to introduce $\tau_1$. Since the essence of the stretched 
exponential function is randomness in the $T_1$ process, 
one may speculate that $T_1(r)$ with a local 
spin density induced by Mg is approximated by a power law, leading 
to the stretched exponential function. In the soliton picture~\cite{Khomskii},
 the soliton which stays in the middle of a segment or near 
the edges due to an interchain coupling, carries spin 1/2, so 
that it can act as an impurity spin.  

\begin{figure}
\epsfxsize=2.8in
\epsfbox{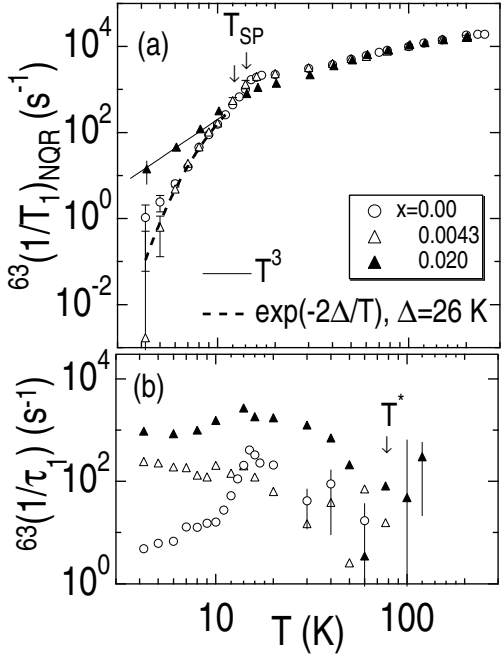}
\vspace{-0.5cm}
\caption{
Log-log plots of $^{63}(1/T_1)_{NQR}$ (a) 
and $^{63}(1/\tau_1$) (b) as functions of temperature 
for undoped $x$=0 and for Mg-doped samples of $x$=0.0043 
and 0.020. In (a), the solid line is a $T^3$ function, 
and the dashed curve is the least-squares fitting result using 
a function of 1/$T_1=R_1$exp$(-2\Delta/T)$.
}
\label{LogT1}
\end{figure}

Figure 3 shows log-log plots of $^{63}(1/T_1)_{NQR}$ (a) and $^{63}(1/\tau_1$) (b) as
functions of temperature for $x$=0, 0.0043 and 0.020. Far above $T^*$, $^{63}(1/T_1)_{NQR}$ for Mg-doped samples
is nearly the same as that for an undoped one. For undoped and $x$=0.0043, the activation-type  temperature
dependence of $^{63}(1/T_1)_{NQR}$ is  observed below $T_{sp}$. The spin gap $\Delta$ is estimated to be
${\sim}$26 K by fits of 1/$T_1$=$R_1$exp(-2$\Delta/T$) ($R_1$ and
$\Delta$ are fitting parameters)~\cite{Ehrenfreund}, which agrees with the value estimated from the
static susceptibility~\cite{Hase}. For $x$=0.020, however, the temperature dependence of $^{63}(1/T_1)_{NQR}$ is
changed into a power-law  type (${\sim}T^3$), probably because of an inhomogeneous 
distribution of $\Delta(r)$. Conventional critical
divergence toward the magnetic ordering does not appear. 

\begin{figure}
\epsfxsize=3.3in
\epsfbox{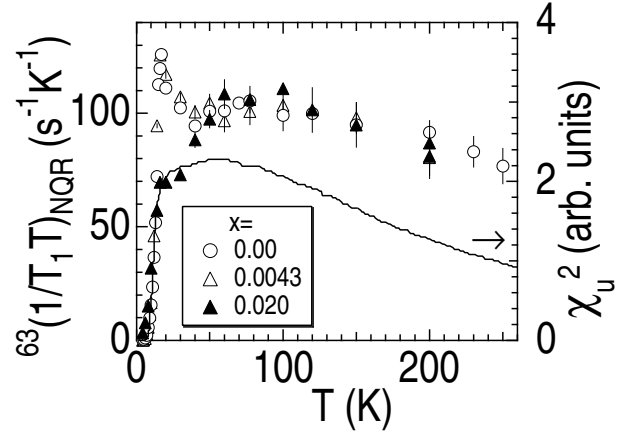}
\vspace{0cm}
\caption{
The temperature and Mg-doping dependence of $^{63}(1/T_1T)_{NQR}$. 
The solid curve is the squared static uniform susceptibility $\chi_u^2$ 
of undoped CuGeO$_3$ reproduced from Ref. [1].
}
\label{W1}
\end{figure} 

Below around $T^*$, $^{63}(1/\tau_1)$ immediately 
increases as the temperature is decreased down to $T_{sp}$ 
even for $x$=0, which indicates the increase of the 
impurity-induced spin correlation. Far below $T_{sp}$, $^{63}(1/\tau_1)$ is systematically enhanced by Mg doping,
which is due to an increase  of the number of impurity relaxation centers. It is likely that 
the origin of $^{63}(1/\tau_1$) for $x$=0 
is due to non-intentionally introduced imperfections (defects, 
dislocations, \dots ). The actual sample is not a perfect crystal, 
because the observed linewidths of $^{63}$Cu NQR spectra of our $x$=0 
($\sim$180 kHz at 4.2 K, $\sim$100 kHz above 100 K) are 
broader than those expected from $T_1$ or $T_2$ broadening (a few kHz), i.e. inhomogeneous broadening. The 
estimated $^{63}\Delta\nu_{1/2}$ for $x$=0 
is nearly the same as or somewhat sharper than the reported values 
below 40 K [16, 17, 22, 27]. In Fig. 3(b), $^{63}\tau_1$ for each $x$ makes a kink at around
$T_{sp}$. In terms of an impurity spin picture, $^{63}(1/\tau_1)$ is nearly proportional to
the life time of the impurity spin  scattered by the host magnetic excitations. Then, the kink of $^{63}(1/\tau_1)$
reflects a change in the host magnetic excitation spectrum at the true transition temperature $T_{sp}$, which is evident 
in $^{63}(1/T_1)_{NQR}$. 

Figure 4 shows the Mg-doping effect on $^{63}(1/T_1T)_{NQR}$. For comparison, the squared static susceptibility
$\chi_u^2$ of undoped CuGeO$_3$ (solid curve) is also reproduced from  Ref. [1]. In general,
1/$T_1T$ is the low frequency  dynamical spin susceptibility at an NQR
frequency summed over  a momentum space via a nuclear-electron coupling [35]. For $x$=0 
above $T_{sp}$, $^{63}(1/T_1T)_{NQR}$ is understood by the sum of the staggered spin susceptibility 
$\chi(q=\pi){\sim}1/T$ and the Bonner-Fisher-type uniform spin susceptibility
$\chi_u$ (to be exact, $\chi_u^2$)~\cite{Mitoh1,Yu}. The upturn of $^{63}(1/T_1T)_{NQR}$ just above 
$T_{sp}$, which is ascribed to the staggered $\chi(q=\pi){\sim}1/T$, is suppressed by Mg doping of $x$=0.020. Then, the contribution 
from the uniform mode $q$=0 is uncovered, being  similar to the temperature dependence of
$\chi_u^2$. The $q=\pi$
mode is easily affected  by imperfection, comparatively more than the $q$=0  mode, as can be seen for La$_2$Cu$_{1-x}$Zn$_x$O$_4$
~\cite{Uchinokura}.  However, one should note that $^{63}(1/T_1T)_{NQR}$
for $x$=0.020 and $\chi_u^2$ for $x$=0 above $T_{sp}$ are similar but do not completely agree 
with each other. The suppression of $^{63}(1/T_1T)_{NQR}$
for $x$=0.020 begins from below 60-120 K more steeply 
than that of $\chi_u^2$ below about 50 K. Thus, the 
further mechanism of the suppression is needed. The precursory 
dimerization enhanced by Mg below around $T^*$ is a 
possible candidate. Our observations of the suppressed ${\chi}(q={\pi}$) and of the deviation between $^{63}(1/T_1T)_{NQR}$ and
$\chi_u^2$ for Mg doped samples will be constraints  on dynamical theory toward the low temperature dimerized antiferromagnetic 
transition. 

To conclude, below $T^*\sim$77 K, the inhomogeneous 
broadening of Cu NQR spectra and the impurity-induced Cu nuclear 
spin-lattice relaxation occur for undoped and more remarkably 
for Mg-doped CuGeO$_{3}$. Precursory dimerization, inhomogeneous 
in real space, is suggested. The host antiferromagnetic correlation 
above $T_{sp}$ is suppressed by Mg doping of $x$=0.020. 
No magnetic critical divergence down to 4.2 K is a puzzle.
 
We thank Dr. J. Kikuchi, and Prof. M. Ogata for stimulating 
discussions. This work was supported by New Energy and Industrial 
Technology Development Organization (NEDO) as Collaborative Research 
and Development of Fundamental Technologies for Superconductivity 
Applications.


\begin{references}
\bibitem{DefT1} In Refs. [16, 17, 22], 1/$T_1$ is estimated by $p(t)=p(0)$exp(-3$t/T_1$). But, under an
asymmetric electric field gradient (an asymmetry factor $\eta\sim$0.16 [16, 17]), all the $x$-, $y$-, and $z$-components of the
local field fluctuations contribute to zero-field Cu $(T_1)_{NQR}$. Then, a numerical factor 3 is not needed.
\end{references}
\end{document}